\documentclass[twocolumn,showpacs,preprintnumbers,amsmath,amssymb,nofootinbib]{revtex4-2}
\usepackage{epsfig}

\usepackage{comment}
\usepackage[table]{xcolor}
\usepackage{graphicx}
\usepackage{dcolumn}
\usepackage{bm}

\usepackage{hyperref}
\hypersetup{colorlinks=true,linkcolor=magenta,anchorcolor=green,citecolor=cyan,filecolor=black,menucolor=black,urlcolor=brown}
\usepackage{slashed}
\newcommand{\ba}{\begin{eqnarray}}
\newcommand{\ea}{\end{eqnarray}}
\newcommand{\be}{\begin{equation}}
\newcommand{\ee}{\end{equation}}

\newcommand{\ecc}{{\mathfrak e}}

\begin{document}

\title{Testing scalar dark matter clumps with Pulsar Timing Arrays }

\author{Philippe Brax and Patrick Valageas}
\affiliation{Universit\'{e} Paris-Saclay, CNRS, CEA, Institut de physique th\'{e}orique, 91191, Gif-sur-Yvette, France}

\begin{abstract}

Scalar dark matter is a viable alternative to particle dark matter models such as
Weakly Interacting Massive Particles (WIMPS). This is particularly the case for scalars
with a low mass $m \gtrsim 10^{-21} {\rm eV}$ as required to make quantum effects macroscopic
on galactic scales.
We point out that by synchronising the measurements of arrival times of pairs of pulsars,
Pulsar Timing Arrays (PTA) could probe ultralight dark matter (ULDM) scenarios with a mass
$10^{-23} {\rm eV}\lesssim m \lesssim 10^{-19} {\rm eV}$ that is greater than the one reached
in standard analysis.
The upper limit on the mass $m$ is set by the time lag $\Delta t$ between the observations
of the two pulsars and could be pushed above $10^{-19}  {\rm eV}$ for $\Delta t$ smaller than
one hour.
However, for these high scalar masses only very high density dark matter clouds could be detected
and the capture rate of neutron stars is too low to provide sufficient statistics.
Significant detection probabilities would thus require direct dark-matter-baryon interactions
that favor the formation of neutron stars within such dark matter clouds, or the discovery of
black hole/pulsar binary systems, taking advantage of the dark matter spike
generated by the black hole.

\end{abstract}
\maketitle

\section{Introduction}

The range of possible masses for the dark matter particles has been largely extended in the last few years \cite{Hui:2021tkt}.
Of particular interest is the low mass region below $m\lesssim 1$ eV where dark matter behaves
like a classical field whose non-relativistic limit is well approximated by a Schr\"odinger equation.
When the mass is low enough, around $m\sim 10^{-22}$ eV, the quantum properties of the dark matter
field at the de Broglie length can reach galactic sizes \cite{Hu2000,Hui:2016ltb}.
Self-interactions alter this picture. In particular quartic self-interactions give rise to a
Gross-Pitaevski equation with cubic terms in the Schr\"odinger equation.
This can change the small distance properties of dark matter as solitons can emerge
that are governed by the balance between gravity and the repulsive self-interactions.
They have a well-defined size $r_a$ related to the self-interaction coupling and the scalar mass
\cite{Chavanis:2011zi,Brax:2019fzb,Garcia:2023abs}.
As such they can be candidates for modifying the small-scale properties of dark matter halos, i.e. they
render them smooth on scales lower than $r_a$.
In addition, they can display peculiar features such a vortex lattices, similar to those found
in laboratory experiments on superfluids \cite{Brax:2025uaw,Brax:2025vdh}.
Self-interacting model can be relevant astrophysically for a larger range of masses than the free fuzzy dark matter case.
Testing scalar dark matter models with larger masses is one of the aims of this article.

The solitons can also have much smaller radii than galactic sizes and represent clumps of dark matter \cite{Kolb:1993zz,Kolb:1993hw}.
In this case, their central density can be much larger than the one averaged on a kpc scale
or the upper bound found in the Earth's environment.
In particular, dense clumps could surround far away objects such as pulsars or binary black holes.
In the latter case, the evolution of the binary system can be affected by the dynamical friction
exerted by the dark matter environment \cite{Boudon:2022dxi,Boudon:2023qbu}.
In addition, the propagation of the gravitational waves emitted by the binary system is affected by
the oscillatory component of the Newtonian potential in the clump, at a pulsation of twice the
scalar mass $\omega=2m$.
This could be detected in the future by LISA or DECIGO if the clumps form around the matter-radiation equality and the dark matter mass is between $10^{-23}$ eV and $10^{-21}$ eV
\cite{Brax:2024yqh,Blas:2024duy}.
This could also be detected in Pulsar Timing Arrays (PTA) through the impact on the propagation
time of the signal from each pulsar \cite{Khmelnitsky:2013lxt}.
This is equivalent to the Sachs-Wolfe effect for the Cosmic Microwave Background (CMB).
Numerical simulations have shown that dark matter solitons are typically embedded within
extended virialized enveloppes that exhibit strong density fluctuations on the de Broglie scale
\cite{Schive2014,Garcia:2023abs}. These fluctuations can be described as quasi-particles
\cite{Hui:2016ltb} and they could also be probed by gravitational waves interferometers
and PTA \cite{Kim:2023pkx,Kim:2023kyy,Eberhardt:2024ocm}.
The PTA results can have important consequences for fundamental physics \cite{Ellis:2023oxs} and in particular  dark matter such as the existence of WIMPs \cite{Gouttenoire:2025wxc} or ultra-light dark matter (ULB) \cite{EPTA:2024gxu,EPTA:2023xxk,Unal:2022ooa,Wu:2024sgk}.
On the other hand, if the dark matter is coupled to the neutrinos its coherent oscillation can impact
the propagation of the neutrinos and be detected in oscillation experiments
\cite{Choi:2019zxy,Cheek:2025kks}.

In this paper we revisit the effect associated with the coherent oscillation of the dark matter
gravitational potential inside solitons, focusing on the correlation between two signals coming from
two pulsars \cite{Cusin:2025xle}.
We point out that in the presence of a dark matter clump around each pulsar, the cross-correlation
exhibits a low-frequency signal, $\omega \ll m$. Moreover, high scalar masses could be probed
if the observations of the two pulsars are synchronised.
We evaluate the signal-to-noise ratio (SNR) when taking into account the white and red noises in the
arrival times. A conservative bound on the SNR allows us to predict that a wider range of scalar masses
can be tested by PTA experiments, with masses up to $m\simeq 10^{-19}$ eV for time-lags of one hour
between the two pulsars.
Interestingly, this signal would also be prominent when the dark matter clumps have a density close
to the one at matter-radiation equality. This is in the same ball-park as the detectability of scalar clumps by LISA or DECIGO. As a result the study of such correlations
in the PTA data would certainly give us clues to what can be expected from future gravitational wave experiments. The synergy between the two types of astronomical observables
is worth pursuing, we intend to come back to it in the future.

The paper is arranged as follows. In a first part \ref{sec:ultra} we recall the essential ingredients
of ultra light dark matter and the corresponding Sachs-Wolfe effect in dark matter clumps.
In section \ref{sec:move}, we introduce the correlation between the arrival times and consider
the SNR for the filtered signal at low frequency.
In section \ref{sec:num}, we derive quantitative results and compare this approach with standard
pulsar time delay analysis.
In App.~\ref{app:red-noise} we derive the contribution of the red noise to the variance of the signal.
In App.~\ref{app:capture} we estimate the encounter and capture probabilities of a neutron star
by a scalar dark matter cloud.

\section{Ultralight dark matter}
\label{sec:ultra}

\subsection{Oscillating scalar field}

We consider ultra-light dark matter scenarios, where a scalar field $\phi(\vec x,t)$ is governed by the
Lagrangian density
\be
{\cal L}_\phi = -\frac{1}{2} (\partial\phi)^2 -  \frac{m^2}{2} \phi^2 - \frac{\lambda_4}{4} \phi^4 ,
\ee
where $\lambda_4$ is the self-interaction coupling constant. These self-interacting scalars could play a role
at short distance on sub-galactic scales and make solitons which could alleviate some of the small-scale dark matter conundra.
In the Newtonian gauge $ds^2 = - (1+2\Phi) dt^2 + (1-2\Psi) d\vec x^{\,2}$ and in the
weak-gravity regime, the scalar field obeys, at leading order, the Klein-Gordon equation
\be
\ddot \phi - \vec\nabla^2 \phi + (1+2\Phi) m^2 \phi + \lambda_4 \phi^3 = 0 .
\label{eq:Klein-Gordon}
\ee
Here we assumed that the self-interactions are small, $\lambda_4 \phi^4 \ll m^2\phi^2$, so that at
zeroth order the scalar field oscillates in the harmonic potential $m^2\phi^2/2$,
with a vanishing pressure $P$ and a constant density $\rho$.
Here $\Phi=\Psi$ is the constant Newtonian potential at leading order.
As pointed out by \cite{Khmelnitsky:2013lxt}, we shall recall below in
Eqs.(\ref{eq:rho-osc})-(\ref{eq:Psi-rho0})
that there are also subleading oscillating components in the density, pressure and
gravitational potential, as well as a small mean pressure,
which is subdominant in the nonrelativistic regime.

In the nonrelativistic regime, it is convenient to introduce a complex scalar field $\psi$ with
\cite{Hui:2016ltb}
\be
\phi = \frac{1}{\sqrt{2 m}} \left( e^{-i m t} \psi + e^{i m t} \psi^\star \right),
\ee
which obeys the Schr\"odinger equation
\be
i \dot\psi = - \frac{\vec \nabla^2 \psi}{2m} + m ( \Phi + \Phi_I ) \psi , \;\;\;
\Phi_I = \frac{3 \lambda_4}{4 m^3} |\psi|^2 .
\ee
This can be mapped to an hydrodynamical picture through the Madelung transform
\cite{Madelung:1927ksh}
\be
\psi = \sqrt{\frac{\rho}{m}} e^{i S} , \;\;\; \vec v = \frac{\vec\nabla S}{m} ,
\ee
which leads to the continuity and Hamilton-Jacobi equations
\be
\dot\rho + \frac{\vec \nabla \cdot ( \rho \vec\nabla S )}{m} = 0 ,
\label{eq:continuity}
\ee
\be
\dot S + \frac{(\vec \nabla S)^2}{2 m} = - m (\Phi_Q + \Phi + \Phi_I ) , \;\;\;
\Phi_Q = - \frac{\vec\nabla^2\sqrt{\rho}}{2m^2\sqrt{\rho}} ,
\label{eq:Hamiton-Jacobi}
\ee
where we introduced the quantum pressure $\Phi_Q$.
Taking the gradient of Eq.(\ref{eq:Hamiton-Jacobi}) gives the Euler equation, with an effective pressure
due to the quantum pressure $\Phi_Q$ and the self-interaction potential $\Phi_I$.

These equations of motion typically lead to the formation of hydrostatic equilibria, also called solitons
or bosons stars, which may be embedded within a larger virialized halo governed by velocity dispersion
\cite{Garcia:2023abs}.
Whereas the soliton is a ground state associated with a vanishing velocity dispersion
and a smooth density profile over a large radius $R_0$, the outer envelope displays large density
fluctuations over the de Broglie wavelength $\lambda_{\rm dB}$.
Typically we consider the regime where the soliton
has a size larger than the de Broglie wavelength and is the result of the equilibrium between the scalar pressure (repulsive) and gravitation.

In this paper, we consider the case where pulsars could be located within dark matter solitons.
Denoting by $\vec v_0$ the collective velocity of the soliton, the hydrostatic equilibrium that
determines the density profile of the soliton reads
$\Phi_Q + \Phi + \Phi_I =$ constant, and as seen from Eq.(\ref{eq:Hamiton-Jacobi}) the phase reads
\be
S = - \mu m t + m \vec v_0 \cdot \vec x - \alpha , \;\;\;
\mu = \frac{v_0^2}{2} + \Phi_Q + \Phi + \Phi_I ,
\label{eq:mu-def}
\ee
where $\alpha$ is a constant phase offset.
Going back to the scalar field $\phi$, this gives
\be
\phi = \frac{\sqrt{2\rho}}{m} \cos( E t + \beta ) , \;\;\;
\beta = \alpha - m \vec v_0 \cdot \vec x ,
\label{eq:phi-osc}
\ee
and
\be
E = m (1 + \mu) = m \left( 1 + \frac{v_0^2}{2} + \Phi_Q + \Phi + \Phi_I \right)  .
\ee
We can check that this is a solution of the Klein-Gordon equation (\ref{eq:Klein-Gordon})
in the nonrelativistic regime, $|\mu| \ll 1$, after averaging over the fast oscillations.

\subsection{Oscillating gravitational potentials}

The oscillating scalar field (\ref{eq:phi-osc}) actually leads to both constant and oscillating components
in the scalar-field energy-momentum tensor $T^\mu_\nu$ \cite{Khmelnitsky:2013lxt},
\be
- T^0_0 = \bar\rho + \rho_{\rm osc} , \;\;\; T^i_i = 3 (\bar P + P_{\rm osc} ) ,
\ee
with at leading order $\bar\rho = \rho$,
\be
\rho_{\rm osc} = - \rho \cos(2\theta) \left( v_0^2 + \bar\Phi_Q + \frac{1}{3} \bar\Phi_I \right)
+ \rho \cos(4\theta) \frac{1}{6} \bar\Phi_I ,
\label{eq:rho-osc}
\ee
where we denote $\theta=Et+\beta$ the argument of the cosine in Eq.(\ref{eq:phi-osc}), and
\be
\bar P = \rho \left( \frac{v_0^2}{3} + \bar\Phi_Q + \frac{\bar\Phi_I}{2} \right) ,
\;\;\;
P_{\rm osc} = - \rho \cos(2\theta) .
\label{eq:P-mean-osc}
\ee
Thus, whereas $\rho_{\rm osc} \ll \bar\rho$ we have $\bar P \ll P_{\rm osc}$ in the
nonrelativistic regime.
From the Einstein equations,
\be
- 2 \vec\nabla^2 \Psi = 8 \pi {\cal G} T_0^0 , \;\;
6 \ddot\Psi - 2 \vec\nabla^2 ( \Psi - \Phi ) = 8 \pi {\cal G} T^i_i ,
\ee
we obtain for the gravitational potential $\Psi$
\be
\nabla^2\bar\Psi = 4\pi {\cal G} \rho_0 , \;\;\; \Psi_{\rm osc} = \frac{\pi {\cal G} \rho_0}{m^2} \cos(2\theta) .
\label{eq:Psi-rho0}
\ee
The second equation gives access to the oscillating part of the Newtonian potential induced by the rapid oscillations of the matter density.

\subsection{Sachs-Wolfe effect and time delays}

As for the Sachs-Wolfe effect for the Cosmic Microwave Background (CMB), when photons travel
from a pulsar towards the Earth, their frequency is modified by the metric fluctuations
as \cite{Khmelnitsky:2013lxt}
\be
\frac{f_e-f_p}{f_p} = \int_{t_p}^{t_e} dt \, \partial_t (\Phi+\Psi) + \Phi_p - \Phi_e .
\ee
Here we denote with the subscript $e$ and $p$ the time and location associated with the Earth and
the pulsar.
The time-independent components $\bar\Psi$ and $\bar\Phi$ of the gravitational potentials
lead to a constant frequency shift that cannot be measured, as we do not know with exact accuracy
the value of the emission frequency $f_p$. Therefore, in the following we focus on the oscillatory components
of the metric potentials.
Writing $\partial_t = \frac{d}{dt} - n_i \partial_i$, where $\vec n$ is the unit vector along the signal
propagation, and integrating, we obtain \cite{Khmelnitsky:2013lxt}
\be
\frac{f_e-f_p}{f_p} = \Psi_e - \Psi_p - \int_{t_p}^{t_e} dt \, n_i \partial_i (\Phi+\Psi) .
\ee
The integrated effect is suppressed by a factor $k/m$ where $k$ is the wave number. Assuming that the gravitational potential is much deeper
in the dark matter cloud around the pulsar than around the Earth, we approximate the frequency shift by
\be
\frac{\delta f}{f}(t) = - \Psi_p \cos( 2 E_p (t-d_p) + 2 \beta_p ) ,
\ee
where we wrote $t_p=t_e-d_p$, with $d_p$ the distance to the pulsar,
and $\Psi_p = \pi {\cal G} \rho_p/m^2$ from Eq.(\ref{eq:Psi-rho0}),
where $\rho_p$ is the density of the dark matter soliton around the pulsar.
As explained above, here we focus on the time-dependent frequency shift.

The frequency shift leads to a time delay $\delta t$ of the pulses measured on the Earth,
\be
\delta t = - \int_0^t dt \frac{\delta f}{f} .
\ee
Keeping again only the time-dependent component, this gives
\be
\delta t = \frac{\Psi_p}{2 m} \sin( 2 E_p t + \gamma_p ) , \;\;\; \gamma_p = - 2 E_p d_p + 2 \beta_p .
\label{eq:delta-t}
\ee

In standard analysis of pulsar timing array data
\cite{Khmelnitsky:2013lxt,Porayko:2018sfa,NANOGrav:2023hvm,EPTA:2024gxu},
one directly uses Eq.(\ref{eq:delta-t})
to search for an ultralight dark matter signal in the set of times of arrival.
In this approach, one approximates $E_p \simeq m$, so that all pulsars may lead to an oscillatory
signal with the same frequency $f = m/\pi$. If the dark matter cloud is very large
($m \lesssim 10^{-22}$ eV) and contains the Earth as well as all the pulsars, one includes the Earth term
and all terms have the same amplitude.
This leads to constraints on ultralight dark matter scenarios with $m \lesssim 10^{-22}$ eV.
This upper limit is set by the typical time interval $\Delta T_{\rm obs}$ between two measurements,
which sets an upper bound on the frequency $f_{\max} = 1/\Delta T_{\rm obs}$ that can be
measured from the data.
With $\Delta T_{\rm obs} \simeq 1$ week, this gives $f_{\max} \sim 2 \times 10^{-6}$ Hz
and $m_{\max}= \pi f_{\max} = 3 \times 10^{-21}$ eV.
A more careful analysis shows that these data provide constraints in the range
$10^{-24}$ eV $\lesssim m \lesssim 10^{-22}$ eV, as one needs at least a few points in a cycle
to  extract  the signal \cite{NANOGrav:2023hvm}.

\section{Moving the observational window to higher scalar mass}
\label{sec:move}

\subsection{Correlation of a pair of pulsars}

In this paper, we investigate whether one can constrain higher dark matter masses $m$ by cross-correlating
the signals from different pulsars.
Indeed, considering two pulsars $a$ and $b$ measured at times $t_{ai}$ and $t_{bj}$, we obtain
\be
\delta t_{ai} \delta t_{bj} =  \frac{\Psi_a \Psi_b}{4 m^2} \sin( 2 E_a t_{ai} + \gamma_a )
\sin( 2 E_b t_{bj} + \gamma_b ) .
\ee
This also reads
\ba
\delta t_{ai} \delta t_{bj} & = & \frac{\Psi_a \Psi_b}{8 m^2} \biggl [
\cos( 2 E_a t_{ai} - 2 E_b t_{bj} + \gamma_a - \gamma_b ) \nonumber \\
&& - \cos( 2 E_a t_{ai} + 2 E_b t_{bj} + \gamma_a + \gamma_b ) \biggl ] .
\label{eq:dta-dtb}
\ea
If $t_{ai}=t_{bj}=t$, the second term oscillates at the angular frequency $\omega \simeq 4 m$,
which is the second harmonic of the one-point signal (\ref{eq:delta-t}), and does not give access
to higher scalar mass $m$.
However, the first term oscillates at the much smaller angular frequency
$\omega = 2 (E_a-E_b) = 2 m (\mu_a-\mu_b) \ll 2 m$.
For a fixed range of frequency probed by an experiment, this could provide constraints on much higher
scalar masses $m$.
Moreover, by filtering the signal as in Eq.(\ref{eq:s-def}) below, one can reach high scalar masses
up to $m \lesssim 1/|\Delta t_{ij}|$ as in Eq.(\ref{eq:regime}) below, where $\Delta t_{ij}$ is the
time-lag between the measurements of the two pulsars.
This upper mass limit could thus be pushed to high values.
The goal of this paper is to investigate these points.

Defining the means and differences
\be
\bar\mu = \frac{\mu_a\!+\!\mu_b}{2} , \; \Delta\mu = \frac{\mu_a\!-\!\mu_b}{2} , \;
\bar\gamma = \frac{\gamma_a\!+\!\gamma_b}{2} , \; \Delta\gamma = \frac{\gamma_a\!-\!\gamma_b}{2} ,
\ee
and
\be
\bar t_{ij} = \frac{t_{ai}+t_{bj}}{2} , \;\; \Delta t_{ij} = \frac{t_{ai}-t_{bj}}{2} ,
\ee
the product (\ref{eq:dta-dtb}) reads
\ba
\delta t_{ai} \delta t_{bj} & = & \frac{\Psi_a \Psi_b}{8 m^2} \big [
\cos( 4 m (1 \! + \! \bar\mu) \Delta t_{ij} \! + \! 4 m \Delta\mu \bar t_{ij} \! + \! 2 \Delta\gamma ) \nonumber \\
&& - \cos( 4 m (1 \! + \! \bar\mu) \bar t_{ij} \! + \! 4 m \Delta\mu \Delta t_{ij}
\! + \! 2\bar\gamma ) \big ] .
\label{eq:dta-dtb-Delta}
\ea
We shall sum over measurements with small time intervals $\Delta t_{ij}$ (i.e., the two pulsars are
observed at two closely separated times) while $\bar t_{ij}$ can span a few years (the duration $T_{\rm obs}$
of the observational campaign).
The second term oscillates with $\bar t_{ij}$ at the fast angular frequency $4m (1+\bar\mu)$
whereas the first term oscillates more slowly at the angular frequency $4m \Delta \mu$.
To distinguish this small oscillatory component in the data, we multiply the time delays by a
similar oscillatory filter and we define the observable
\be
s = \frac{1}{N_{ij}} \sum_{ij} \delta t_{ai} \delta t_{bj} \cos(4 \omega \bar t_{ij} ) ,
\label{eq:s-def}
\ee
where we sum over a set of $N_{ij}$ pairs of times of arrival $\{t_{ai},t_{bj}\}$
and $\omega > 0$. In the regime
\be
4 m T_{\rm obs} \gg \pi , \;\; 4 m |\Delta t_{ij}| \ll \pi , \;\;
\left| m | \Delta\mu | - \omega \right| T_{\rm obs} \ll \frac{\pi}{8} ,
\label{eq:regime}
\ee
the second term in (\ref{eq:dta-dtb-Delta}) shows many oscillations that cancel out while the first term
gives a dominant contribution of the form
\be
s_{\rm DM} = \frac{\Psi_a \Psi_b}{16 m^2} \cos(2 \Delta \gamma) ,
\label{eq:s-DM}
\ee
associated with a dark matter cloud around each pulsar.

\subsection{Dark matter mass window}

The regime associated with the conditions (\ref{eq:regime}) corresponds to the mass window
\ba
&& m \gg \left( \frac{T_{\rm obs}}{1 \, {\rm yr}} \right)^{-1} 2 \times 10^{-23} \, {\rm eV} , \nonumber \\
&& m \ll \left( \frac{| \Delta t |}{1 \, {\rm hour}} \right)^{-1} 10^{-19} \, {\rm eV} , \nonumber \\
&& \left | m | \Delta\mu | - \omega \right | \ll \left( \frac{T_{\rm obs}}{1 \, {\rm yr}} \right)^{-1}
8 \times 10^{-24} \, {\rm eV} .
\label{eq:regime-num}
\ea
We can see that for $| \mu | \lesssim 10^{-3}$ the width over $m$
of a probe at frequency $\omega$ is not too narrow, which makes such an analysis possible.
Depending on the total observational time $T_{\rm obs}$ and the time difference $\Delta t$
between the measurement times of two pulsars, one may probe in this fashion dark matter scenarios
with masses in the range $10^{-23}$ eV $< m < 10^{-19}$ eV. Of course this assumes that the cadence of pulsar observations could
be as high as one per hour. This has to be adapted to future forecast for such observations.
Thus, our filtering method  could give access to scalar masses somewhat above the standard approach, which typically
probes $m \lesssim 10^{-22}$ eV.
In particular, the upper bound on the scalar mass $m$ that can be reached is set by the
time lag $\Delta t$ between the measurements of the two pulsars.
The main point of this paper is thus that by making this time very small, for instance
by observing the two pulsars on an overlapping time interval, one could have access to high $m$.
On the other hand, to be detectable large $m$ models would require high-density dark matter clouds,
because of the $1/m^2$ factors in Eqs.(\ref{eq:Psi-rho0}) and (\ref{eq:s-DM}).

The Compton wavelength reads
\be
\lambda_C = \frac{2\pi}{m} = \left( \frac{m}{10^{-20} \, {\rm eV}} \right)^{-1} 4 \times 10^{-3} \, {\rm pc} .
\ee
Therefore, the dark matter clouds that could be probed by such analysis would typically have sizes
above $0.01$ pc, as $R > \lambda_C$. As a result, we do not consider highly dense clumps with large Newtonian potentials.

\subsection{Noise contributions}

\subsubsection{Noise contributions to the timing residuals}

As in standard analysis of pulsar times of arrival
\cite{NANOGrav:2023hvm},
we write the timing residuals (with respect to a specific model of the pulsar
and of the motion of the Earth) as
\be
{\bf \delta t} = {\bf M} \cdot {\bf\epsilon} + {\bf w} + {\bf r} + {\bf \delta t}_{\rm GW} + {\bf \delta t}_{\rm DM} ,
\ee
where ${\bf \delta t}$ is the vector of the timing residuals, $\{ \delta t_{ai} , \delta t_{bj} \}$.
The first term on the right-hand side arises from the error ${\bf\epsilon}$ on the parameters
of the underlying deterministic timing model, using a linear approximation appropriate for small perturbations.
In the following we assume that the model parameters ${\bf\epsilon}$ have already been calibrated
from a standard analysis of the pulsar data and we discard this term.
The second term is a white noise that is left after subtracting known systematics.
The third term, often denoted ${\bf F} \cdot {\bf a}$, is a red noise component.
The fourth term corresponds to the stochastic gravitational wave background.
The fifth term is the time delay due to the dark matter cloud, given by Eq.(\ref{eq:delta-t})
in our case.

We assume that the noises associated with different pulsars are uncorrelated and we write
\be
\langle w_{ai} w_{bj} \rangle = \delta_{ab} \delta_{ij} \sigma_{ai}^2 , \;\;\;
\langle r_{ai} r_{bj} \rangle = \delta_{ab} C^r_a(t_i-t_j) ,
\label{eq:noise-a-b}
\ee
where $\langle\dots\rangle$ is the average over the noise, which we assume to be Gaussian with
zero mean, and $C^r_a$ is the correlation function of the red noise for pulsar $a$.

We also write the stochastic gravitational wave background as a Gaussian noise of zero mean,
with a pulsar correlation $\Gamma_{ab}$ given by the Hellings $\&$ Downs overlap reduction function \cite{Hellings:1983fr},
\be
\langle \delta t_{{\rm GW} ai} \delta t_{{\rm GW} bj} \rangle = \Gamma_{ab} C_{\rm GW}(t_{ai}-t_{bj}) .
\ee

We define the power spectra of the red noise and of the stochastic gravitational wave background
as
\be
C^r_a(t) = \int_{f_{\min}}^{f_{\max}} df \, \cos( 2\pi f t ) \, P_a(f) ,
\label{eq:Ca-def}
\ee
and
\be
C_{\rm GW}(t) = \int_{f_{\min}}^{f_{\max}} df \, \cos( 2\pi f t ) \, P_{\rm GW}(f) ,
\label{eq:C-GW-def}
\ee
assuming most of their contribution comes from a finite range of frequencies.

\subsubsection{Statistics of measurement times}

Hereafter, we assume the following observational strategy. For each measurement at time $t_{ai}$
of the pulsar $a$ we associate a single measurement time $t_{bj}$ for the pulsar $b$
(the closest available time). Thus, the observable (\ref{eq:s-def}) reads as a single sum
over $N$ measurement time pairs,
\be
s = \frac{1}{N} \sum_{i=1}^N \delta t_{ai} \delta t_{bi} \cos(4 \omega T_i ) , \;\; T_i = \frac{t_{ai}+t_{bi}}{2} .
\ee
We assume that the measurement times are roughly equally spaced over the total
observational time $T_{\rm obs}$,
\be
t_{ai} = i \Delta T + \Delta t_{ai} , \;\; t_{bi} = i \Delta T + \Delta t_{bi} , \;\; \Delta T = \frac{T_{\rm obs}}{N} ,
\label{eq:Delta-t_ai-Delta-t_bi-def}
\ee
where $\Delta t_{ai}$ and $\Delta t_{bi}$ are independent Gaussian variables of variance
$\sigma_t \ll \Delta T$.
We denote with the angular brackets the average over the noise, as in Eq.(\ref{eq:noise-a-b}),
and with an overbar the average over the measurement times $\Delta t_{ai}$ and $\Delta t_{bi}$,
as in Eq.(\ref{eq:s-mean-split}) below.
This statistical analysis allows us to estimate the typical signal-to-noise ratio as a function
of the main properties of the measurement campaign: the total observational time $T_{\rm obs}$,
the rough periodicity $\Delta T$ of the measurements, and the irregularity $\sigma_t$ of the
observations due to various technical constraints.

\subsubsection{Mean signal}

As the noise terms of the two pulsars are uncorrelated, as in (\ref{eq:noise-a-b}),
the average of the observable $s$ over the noise is given by the contributions from the
dark matter oscillations and from the stochastic gravitational wave background,
\be
\overline{ \langle s \rangle } = s_{\rm DM} + s_{\rm GW} .
\label{eq:s-mean-split}
\ee
The dark matter contribution (\ref{eq:s-DM}) reads
\be
s_{\rm DM} = \frac{\Psi_a \Psi_b}{16 m^2} \cos(2 \Delta \gamma)
= \frac{\pi^2 {\cal G}^2 \rho_a \rho_b}{16 m^6}  \cos(2 \Delta \gamma)  ,
\label{eq:s-bar-DM}
\ee
which depends crucially on the matter density in the clumps.

The stochastic gravitational wave background contribution reads
\ba
s_{\rm GW} & = & \Gamma_{ab} \cos[ 2 \omega (N+1) \Delta T ] e^{-2 \sigma^2 \omega^2}
\frac{\sin(2 \omega T_{\rm obs})}{N \sin(2\omega \Delta T)} \nonumber \\
&& \times \int_{f_{\min}}^{f_{\max}} df \, P_{\rm GW}(f) e^{-2\pi^2 \sigma^2 f^2} ,
\label{eq:s-bar-GW}
\ea
where we introduced $\sigma^2 = 2 \sigma_t^2$.
This contribution is suppressed by the trigonometric factors for $\omega T_{\rm obs} \gg \pi$.
In the following, we assume that this contribution can be distinguished from the dark matter
signal (\ref{eq:s-bar-DM}) thanks to its specific angular correlation $\Gamma_{ab}$ and its
dependence on $\omega$.

\subsubsection{Variance of the signal}

\paragraph{White noise}

The contribution of the white noise to the variance of the observable $s$ reads
\be
\langle s^2 \rangle_w = \frac{1}{N^2} \sum_i \cos^2(4\omega T_i) \sigma_{ai}^2
\sigma_{bi}^2 .
\ee
Taking $\sigma_{ai}= \sigma_a$ and $\sigma_{bi}=\sigma_b$ and averaging over the measurement
times, we obtain for $\sigma_w^2 = \overline{\langle s^2 \rangle_w}$ the two regimes
\ba
&& \omega T_{\rm obs} \ll \pi : \;\; \sigma_w^2 =  \frac{\sigma_a^2 \sigma_b^2}{N} , \nonumber \\
&& \omega T_{\rm obs} \gg \pi : \;\; \sigma_w^2 =  \frac{\sigma_a^2 \sigma_b^2}{2N} .
\label{eq:sigma-w}
\ea
They only differ by a numerical factor coming from the averaging of $\cos^2$.

\paragraph{Red noise}

The contribution of the red noise reads
\ba
\langle s^2 \rangle_r & = & \frac{1}{2 N^2} \sum_{i,i'} \left[ \cos( 4 \omega (T_i \! - \! T_{i'}) ) +
\cos( 4 \omega (T_i \! + \! T_{i'}) ) \right] \nonumber \\
&& \times C_a^r(t_{ai} - t_{ai'}) C_b^r(t_{bi}-t_{bi'}) .
\label{eq:s2r-Ca-Cb}
\ea
Using Eq.(\ref{eq:Ca-def}), in the regime
\be
\omega \ll \frac{\pi}{T_{\rm obs}} , \;\; f_{\max} \ll \frac{1}{\Delta T} , \;\; N \gg 1 ,
\label{eq:sigma2r-regime1}
\ee
the variance due to the red noise becomes
\be
\sigma_r^2 = \overline{\langle s^2 \rangle_r} = \frac{1}{2 T_{\rm obs}} \int_{f_{\min}}^{f_{\max}} df \,
P_a(f) P_b(f) ,
\label{eq:sigma2-r-1}
\ee
whereas in the regime
\be
\frac{\pi}{T_{\rm obs}} \ll \omega \ll f_{\min} , \;\;
f_{\max} \ll \frac{1}{\Delta T} , \;\; N \gg 1 ,
\label{eq:sigma2r-regime2}
\ee
it reads
\be
\sigma_r^2 = \frac{1}{4 T_{\rm obs}} \int_{f_{\min}}^{f_{\max}} df \, P_a(f) P_b(f) .
\label{eq:sigma2-r-2}
\ee
We give in App.~\ref{app:red-noise} more details on the derivation of Eqs.(\ref{eq:sigma2-r-1}) and (\ref{eq:sigma2-r-2}).

\paragraph{Stochastic gravitational wave background}

The computation of the variance due to the stochastic gravitational wave background is similar
to that of the red noise. Using $\Gamma_{aa}=1$,
in the regime (\ref{eq:sigma2r-regime1}) we obtain
\be
\sigma_{\rm GW}^2 = \frac{1+ \Gamma_{ab}^2}{2 T_{\rm obs}} \int_{f_{\min}}^{f_{\max}} df \,
P_{\rm GW}(f)^2 ,
\label{eq:sigma2-GW-1}
\ee
while in the regime (\ref{eq:sigma2r-regime2}) we obtain
\be
\sigma_{\rm GW}^2 = \frac{1+ \Gamma_{ab}^2}{4 T_{\rm obs}} \int_{f_{\min}}^{f_{\max}} df \,
P_{\rm GW}(f)^2 .
\label{eq:sigma2-GW-2}
\ee

\section{Numerical results}
\label{sec:num}

\subsection{Signal-to-noise ratio}

\begin{figure}
\centering
\includegraphics[height=7cm,width=0.5\textwidth]{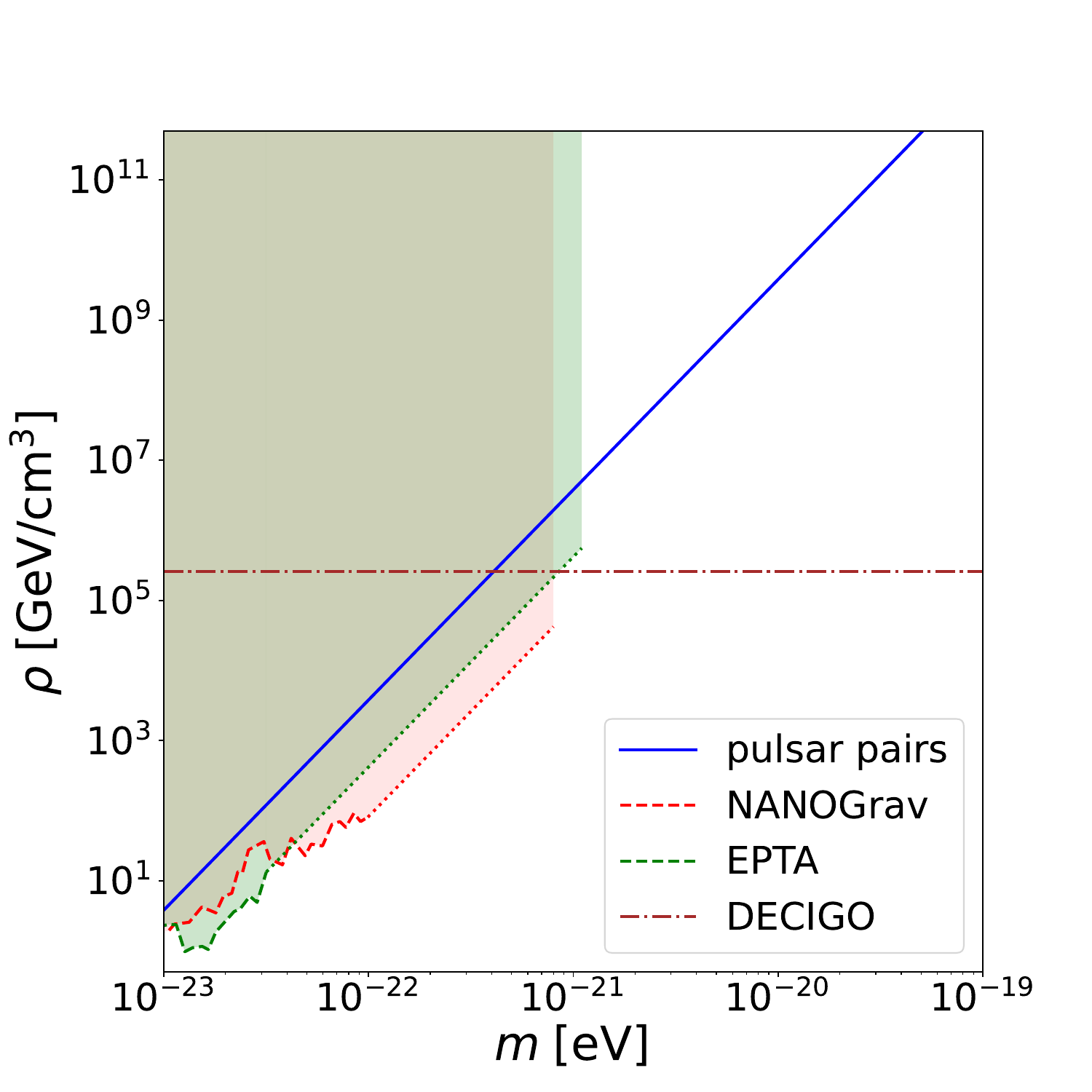}
\caption{
The shaded regions are the exclusion domains obtained from standard
pulsar time delays analysis \cite{NANOGrav:2023hvm,EuropeanPulsarTimingArray:2023egv}.
The blue solid line is our estimate (\ref{eq:rho-SNR}).
The horizontal dot-dashed line is the detection limit of a dark matter cloud, which would contain
a white-dwarf binary system, for the observation of the waveform by the future satellite DECIGO.
}
\label{fig:rho}
\end{figure}

We define the signal-to-noise ratio ${\rm SNR}$ as
\be
{\rm SNR} = \left | \frac{\langle s \rangle}{\sigma_s} \right | ,
\ee
where $\sigma_s$ is the variance due to the white and red noises, which we add in quadrature,
\be
\sigma_s^2 = \sigma_w^2 + \sigma_r^2 + \sigma_{\rm GW}^2 .
\ee
From Eqs.(\ref{eq:sigma-w}), (\ref{eq:sigma2-r-2}) and (\ref{eq:sigma2-GW-2}),
pulsar timing arrays can currently reach
$\sqrt{\sigma_s} \sim 10^{-7}$ s \cite{NANOGrav:2023ctt}.
The gravitational wave background leads to a floor for the variance $\sigma_s^2$
that can only be reduced by increasing the total observational time $T_{\rm obs}$ or
by a thorough analysis of the signal taking into account its detailed spatial
and temporal characteristics, in order to discriminate the dark matter signal and remove
degeneracies.
This goes beyond the signal-to-noise ratio estimate that we consider in this paper
and is left for future work.
A signal-to-noise ratio above unity corresponds to
\be
{\rm SNR} > 1 : \;\; \frac{\sqrt{\rho_a \rho_b}}{m^3} > \frac{4 \sqrt{\sigma_s}}{\pi {\cal G} \sqrt{|\cos(2\Delta\gamma)|}} .
\ee
Taking for simplicity $\rho_a=\rho_b=\rho$ and $|\cos(2\Delta\gamma)|=1/2$,
this gives the detection criterion
\be
\rho_{\rm SNR} > \frac{\sqrt{\sigma_s}}{10^{-7} \, {\rm s}} \left( \frac{m}{10^{-20} \, {\rm eV}} \right)^3
\, 5 \times 10^9 \, {\rm GeV/cm^3} ,
\label{eq:rho-SNR}
\ee
which also reads
\be
\rho_{\rm SNR} > \frac{\sqrt{\sigma_s}}{10^{-7} \, {\rm s}} \left( \frac{m}{10^{-20} \, {\rm eV}} \right)^3
3 \times 10^{15} \, \bar\rho_0 ,
\label{eq:rho-SNR-rho0}
\ee
where $\bar\rho_0$ is the mean cosmological matter density today.

We show in Fig.~\ref{fig:rho} our result (\ref{eq:rho-SNR}), which we compare with the
bounds on ultralight dark matter densities already obtained by standard pulsar time delay analysis
\cite{NANOGrav:2023hvm,EuropeanPulsarTimingArray:2023egv}.
We extend the bounds published in these papers to higher masses, up to $m = \pi / \Delta t$ with
$\Delta t = 1$ month for \cite{NANOGrav:2023hvm} and
$\Delta t = 3$ weeks for \cite{EuropeanPulsarTimingArray:2023egv}, with a slope $\rho \propto m^3$
which should hold for $m \gtrsim 1/T_{\rm obs}$ \cite{NANOGrav:2023hvm}.
We also display the bounds on the dark matter environment that could be obtained from the
observation of gravitational waves emitted by a white-dwarf binary system by DECIGO
\cite{Kawamura:2020pcg,Kinugawa2022} following
from the impact of the dynamical friction within the dark matter cloud on the waveform
\cite{Brax:2024yqh,Boudon:2023vzl}.
We can see that our method, based on the correlation of a pair of pulsars, is less efficient than the
standard PTA analysis by about one to two orders of magnitude. This could be expected, as our bound
corresponds to only one pair of pulsars and the analysis (\ref{eq:rho-SNR}) is not necessarily optimal.
A refined data analysis, with Markov-chain Monte Carlo samplers and several pulsar pairs,
may somewhat improve our bound.
The advantage of our approach is rather to enlarge the parameter space to higher scalar mass $m$.
On the other hand, we can see that at these larger masses, $m \gtrsim 10^{-21}$ eV, other probes
may be more competitive, such as the gravitational waves emitted by binary systems
(through the impact of dynamical friction on the waveform).

We can also see that for $m \gtrsim 10^{-21}$ eV, where our new  method could go beyond standard
analysis, only very dense clouds could be detected, as also shown in Eq.(\ref{eq:rho-SNR}).
The densities found in Fig.~\ref{fig:rho} can be compared with the mean dark matter density
around the solar system in the Milky Way, $\bar \rho_{\rm MW} = 0.4$ Gev/cm$^3$.
They correspond to dark matter clouds that would be denser by a factor $10^6$ at least,
associated with a highly inhomogeneous dark matter density field.

\subsection{Self-consistency checks}

As seen in Eq.(\ref{eq:Psi-rho0}), the amplitude of the oscillating gravitational potential $\Psi_{\rm osc}$
is related to the density by $| \Psi_{\rm osc} | = \pi {\cal G} \rho/m^2$, which reads
\ba
\rho_{\Psi_{\rm osc}} & = & \frac{ | \Psi_{\rm osc} | }{0.1} \left( \frac{m}{10^{-20} \, {\rm eV}} \right)^2
6 \times 10^{19} \, {\rm GeV/cm^3} \nonumber \\
& = & \frac{ | \Psi_{\rm osc} | }{0.1} \left( \frac{m}{10^{-20} \, {\rm eV}} \right)^2 4 \times 10^{25}
\, \bar\rho_0 .
\label{eq:rho-Psi-osc-bound}
\ea

The static gravitational potential is related to the density by $\bar\Psi \sim {\cal G} \rho R^2$,
where $R$ is the radius of the dark matter cloud.
The latter must be larger than the Compton wavelength, $\lambda_C= 2 \pi/m$.
This gives the upper bound
\ba
\rho_{\bar\Psi} & < & \frac{ | \bar\Psi | }{0.1} \left( \frac{m}{10^{-20} \, {\rm eV}} \right)^2
5 \times 10^{18} \, {\rm GeV/cm^3} , \nonumber \\
\rho_{\bar\Psi} & < & \frac{ | \bar\Psi | }{0.1} \left( \frac{m}{10^{-20} \, {\rm eV}} \right)^2 3
\times 10^{24} \, \bar\rho_0 ,
\label{eq:rho-bar-Psi-bound}
\ea
which is somewhat stronger than (\ref{eq:rho-Psi-osc-bound}).
The fact that the signal-to-noise criterion (\ref{eq:rho-SNR-rho0}) is much below the upper bounds
(\ref{eq:rho-Psi-osc-bound}) and (\ref{eq:rho-bar-Psi-bound}) shows that the high densities
required for a detection are not excluded by self-consistency arguments.

\subsection{Representative example}

Taking as a representative case at the detection limit (\ref{eq:rho-SNR}),
\be
m= 10^{-20} \, {\rm eV}, \;\;\; \rho = 5 \times 10^9 \, {\rm GeV/cm^3} ,
\label{eq:ex-rho}
\ee
and choosing a radius $R = 0.1$ pc, we obtain
\be
R = 0.1 \, {\rm pc} , \;\; \bar\Psi = 6 \times 10^{-8} , \;\; \Psi_{\rm osc} = 8 \times 10^{-12} ,
\label{eq:ex-R-Psi_Psi_osc}
\ee
while the mass of the dark matter soliton is
\be
M = 5.5 \times 10^5 \, M_\odot .
\label{eq:ex-M}
\ee
Dark matter clouds with such a large density, $\rho = 3 \times 10^{15} \bar\rho_0$, must have
formed at a redshift $z \sim 10^5$, somewhat before the matter-radiation equality.
Although the clumps have relatively large masses, the fact that their Newtonian potential is so low implies
that their effects on strong gravitational lensing will be small.

We estimate in App.~\ref{app:capture} the encounter and capture probabilities of a neutron star
with dark matter clouds as in Eqs.(\ref{eq:ex-R-Psi_Psi_osc})-(\ref{eq:ex-M}).
We find that the probability of encounter of a neutron star with such a cloud, over one Hubble time,
is rather low, $P_{\rm enc} \simeq 0.003$ from Eq.(\ref{eq:P-enc}).
The probability of capture, through the loss of energy during the crossing of the cloud by
dynamical friction, is even lower, $P_{\rm cap} \simeq 4\times 10^{-8}$ from
Eq.(\ref{eq:P-cap}).
This result happens to be independent of the cloud properties (within its regime of validity)
and is thus quite general.

Therefore, to have a chance of observing a pulsar inside such scalar dark matter clouds
we would need at least 300 pulsars and a $100\%$ capture efficiency, relying for instance
on direct dark-matter-baryons interactions.
A second and more promising scenario may be to increase the star formation rate inside
such dark matter clouds.
Then, a larger fraction of pulsars could be embedded within such clouds, simply because they were
born inside rather than being captured later along their orbit in the Milky Way.
Finally, a third case would be the observation of black hole/pulsar binary systems
\cite{Shao:2018qpt}.
The black hole could generate a dark matter spike with a large density
\cite{Brax:2019npi} (with a value that typically depends on the sign and amplitude of the dark
matter self-interactions), which could be detected
if the pulsar orbit is close enough to the black hole.

With $m | \Delta\mu | \sim m | \bar\Psi | \sim 10^{-26}$ eV, we can see that the third condition
in (\ref{eq:regime-num}) is simply an upper bound on $\omega$.  Here we have taken $\Delta \mu=10^{-6}$,
as from Eq.(\ref{eq:mu-def}) we typically have $\mu \sim \max( | \bar\Phi | , v_0^2 )$, and in the case
(\ref{eq:ex-R-Psi_Psi_osc}) we have $\bar\Phi = \bar\Psi \sim 10^{-7}$ whereas we expect the clump
velocities to be of the order of the typical galactic halo velocity $v_0 \sim 10^{-3}$.
Therefore, such dark matter clouds could be probed by simply taking $\omega=0$ in the definition
(\ref{eq:s-def}) of the observable $s$.
We hope to come back to optimising this new technique
for the extraction of new physics from the PTA signal in the near future.

\section{Conclusion}

We have considered the effects of dense ultra-light dark matter clumps surrounding pulsars.
The Pulsar Timing Array experiments can be used as probes of such dark matter scenarios.
Indeed, in these models solitons can form with a coherent oscillation of the underlying dark matter
scalar field. This leads to a subleading oscillatory component for the local gravitational potential.
These oscillations are directly transcribed via the Sachs-Wolfe effect to the arrival times of pulsar
signals.
As a result, the correlation between pulsar signals from different regions of space will show
oscillations of two types, if these pulsars are embedded in such dark matter clumps.
There is a fast oscillation coming from the coherent behaviour of the dark matter clouds at pulsation
$\omega \simeq 4m$ and a lower frequency signal $\omega = 4 m \Delta \mu$ arising from a beating
mode due to the correlation between two pulsar signals.
This is because the pulsation of the dark matter field is not exactly $m$ but $E=m(1+\mu)$
with $\mu = v^2/2+\Phi_{\rm tot}$ in the nonrelativistic regime.
The fast component at $\omega \simeq 4m$ has already been studied in detail
\cite{Khmelnitsky:2013lxt,NANOGrav:2023hvm}.
In this paper we have pointed out that the slower component could allow PTAs to probe higher
scalar masses.

Filtering this lower frequency component gives direct access to local properties of the clumps and is
sensitive to the mass $m$ of the scalars.
Typically this pulsation is of the order of $4 m \times \max(v^2/2,| \bar\Phi |)$, where $\bar\Phi$
is the gravitational potential of the clump and $v$ its velocity, which we expect to be of the order
of the rotational $v_0 \sim 10^{-3}$ in the Milky Way.
We find that extracting this slow varying correlation could give access to masses for the scalar
as high as $10^{-19}$ eV, if the time lag between the measurements of the two pulsars is of the order
of one hour.
If the measurements are synchronised, e.g. the pulsars are observed on overlapping time intervals,
arbitrarily high dark matter masses can be probed.
However, higher scalar masses require denser clumps to be detectable.

The associated clumps would be relatively large with sizes of a fraction of parsec and large masses.
On the other hand as their Newtonian potential is small, they would appear as soft objects rather than
compact ones, thus evading strong lensing constraints.
As their density is typically larger than the one of matter-radiation equality, the observation
of the slow frequency signal in the correlation between pulsar arrival times would give us a probe
of the late radiation era.
Surprisingly, the same regime of masses and densities would also be probed by the crossing of dark
matter clumps by black hole or white dwarf binaries, as will be observed by LISA or DECIGO
\cite{Boudon:2023vzl,Brax:2024yqh}.
As the PTA data are currently available, we intend to come back to the study of this range of masses
$10^{-23}\ {\rm eV}\lesssim m \lesssim 10^{-19}\ {\rm eV}$ where the PTA signal at low frequency
could be a probe of such ultra light scalar dark matter scenarios.

Unfortunately, the capture rate of neutron stars by such scalar dark matter clumps is very low.
The encounter probability of a neutron star with a dark matter cloud is below $1\%$ and
the capture probability, through dissipation by dynamical friction, is further reduced below
$10^{-7}$.
Therefore, such dark matter clouds could only be detected by the method presented in this paper
if the efficiency of the capture process reaches $100\%$ (e.g., through direct dark-matter-baryon
interactions) or if (neutron) stars are preferentially born within such clouds.
An alternative is to look for black hole/pulsar binary systems, where the dark matter spike
generated by the accretion onto the black hole can reach high densities.

\appendix

\section{Red noise}
\label{app:red-noise}

In this appendix we detail the derivation of Eqs.(\ref{eq:sigma2-r-1}) and (\ref{eq:sigma2-r-2}).

In the regime (\ref{eq:sigma2r-regime1}) we can write Eq.(\ref{eq:s2r-Ca-Cb}) as
\ba
\langle s^2 \rangle_r & = & \frac{C_a(0) C_b(0)}{N} + \frac{1}{N^2} \sum_{i \neq i'} \int df \, df' \,
P_a(f) P_b(f') \;\;\; \nonumber \\\
&& \times \cos[2\pi f(t_{ai}-t_{ai'})] \cos[2\pi f' (t_{bi}-t_{bi'})] ,
\ea
where we used Eq.(\ref{eq:Ca-def}).
Using
\be
\int_{-\infty}^{\infty} \frac{dt}{\sqrt{2\pi} \sigma} e^{-t^2/(2\sigma^2)} \cos[a (t+b)] =
e^{-a^2 \sigma^2/2} \cos(a b) ,
\ee
we can integrate over the probability distribution of the Gaussian variables
$\Delta t_{ai}$ and $\Delta t_{bi}$, defined in (\ref{eq:Delta-t_ai-Delta-t_bi-def}),
\ba
&& \hspace{-0.2cm} \sigma_r^2 = \frac{C_a(0) C_b(0)}{N} + \frac{1}{N^2} \sum_{i \neq i'} \int df \, df' \,
P_a(f) P_b(f') \nonumber \\\
&& \hspace{-0.2cm} \times e^{-2 \pi^2 \sigma^2 (f^2+f'^2)} \cos[2\pi f(i-i') \Delta T] \cos[2\pi f' (i-i') \Delta T] ,
\nonumber \\
&&
\ea
where we again defined $\sigma^2 = 2 \sigma_t^2$.
Writing the product of cosines as a sum of two cosines and using
\be
\sum_{i,i' =1}^N \cos[(i-i')\theta] = \left[ \frac{\sin(N\theta/2)}{\sin(\theta/2)} \right]^2
\to N \pi \delta_D(\theta/2) ,
\label{eq:sum-i-i'-Dirac}
\ee
for $N \to \infty$ and $|\theta| < \pi$,
we obtain in the regime (\ref{eq:sigma2r-regime1}), where $f\sigma \ll 1$,
the expression (\ref{eq:sigma2-r-1}).

In the regime (\ref{eq:sigma2r-regime2}), we keep the cosine terms $\cos(4 \omega T)$ in
Eq.(\ref{eq:s2r-Ca-Cb}), which we write as products of cosines and sines of
$2\omega(t_{ai}-t_{ai'})$ and $2\omega(t_{bi}-t_{bi'})$.
Then, using
\ba
&& \int_{-\infty}^{\infty} \frac{dt}{\sqrt{2\pi} \sigma} e^{-t^2/(2\sigma^2)} \cos[a (t+b)]
\cos[ a' (t+b)] = \nonumber \\
&& \frac{1}{2} e^{-(a+a')^2 \sigma^2/2} \biggl [ e^{2 a a' \sigma^2}
\cos[ (a-a') b] + \cos[ (a+a') b] \biggl ] , \nonumber \\
\ea
and
\ba
&& \int_{-\infty}^{\infty} \frac{dt}{\sqrt{2\pi} \sigma} e^{-t^2/(2\sigma^2)} \sin[a (t+b)]
\cos[ a' (t+b)] = \nonumber \\
&& \frac{1}{2} e^{-(a+a')^2 \sigma^2/2} \biggl [ e^{2 a a' \sigma^2}
\sin[ (a-a') b] + \sin[ (a+a') b] \biggl ] , \nonumber \\
\ea
we can integrate over the Gaussian variables $\Delta t_{ai}$ and $\Delta t_{bi}$.
Expanding again products of cosines and sines and using Eq.(\ref{eq:sum-i-i'-Dirac}),
we obtain the expression (\ref{eq:sigma2-r-2}) in the regime (\ref{eq:sigma2r-regime2}).

\section{Capture rate}
\label{app:capture}

In this appendix we estimate the capture rate of a star by a scalar dark matter cloud.
A rather similar analysis was presented in \cite{Esser:2025kua} for the capture of stars by
the dark matter spike around primordial black holes.
To be captured by the dark matter cloud, we consider that a star coming from a large distance
with the relative velocity $v_\infty$ and the impact parameter $b$ must first enter the cloud and second
dissipate enough energy $\Delta E$ by dynamical friction so that it moves from a free to a bound state.

\subsection{Encounter rate}

The number of encounters per unit time $d\Gamma$ of a star with dark matter clouds of relative velocity
$v_\infty$ and impact parameter $b$ is given by
\be
d\Gamma_{\rm enc} = n v_\infty f(v_\infty) dv_\infty 2 \pi b db ,
\ee
where $n$ is the number density of dark matter clouds while $f(v_\infty)$ is the distribution function
of the relative velocity, normalized to unity.
The local dark matter density in the Milky Way (along the stellar orbit), is
\be
\bar\rho_{\rm MW} = n M , \;\;\ M = \frac{4\pi}{3} \rho R^3 ,
\ee
where $\rho, M$ and $R$ are the density, mass and radius of a dark matter cloud
(we neglect their dispersion and assume a sharply peaked dark matter cloud distribution).
This gives
\be
d\Gamma_{\rm enc} = \frac{3 \bar\rho_{\rm MW}}{2\rho R^3} v_\infty f(v_\infty) dv_\infty b db .
\label{eq:dGamma}
\ee
A first condition for the capture of a distant star is that it enters the cloud. Considering a hyperbolic
orbit of impact parameter $b$ and relative velocity $v_\infty$ at infinity, the eccentricity $\ecc$
is given by (see for instance \cite{poisson_will_2014} for the properties of Keplerian orbits)
\be
\ecc^2 = 1 + \frac{4 v_\infty^4 b^2}{v_{\rm esc}^4 R^2} , \;\; \mbox{with} \;\;
v_{\rm esc}^2 = \frac{2{\cal G} M}{R} ,
\ee
where we introduced the escape velocity $v_{\rm esc}$ from the cloud.
For dark matter clouds such as (\ref{eq:ex-R-Psi_Psi_osc})-(\ref{eq:ex-M}) we have
\be
R=0.1 \, {\rm pc} , \;\; M= 5.5 \times 10^5 M_\odot : \;\;v_{\rm esc} = 217 \, {\rm km/s} .
\label{ex:v-esc}
\ee
As we shall see below, the relative velocity $v_\infty$ will be much smaller than the escape velocity,
$v_\infty \ll v_{\rm esc}$, so that $\ecc \simeq 1$ (i.e., the star is close to the limit between
free and bound orbits).
Then, $r_{\rm peri} \simeq p/2$, where $r_{\rm peri}$ is the perihelion and $p$ the semi-latus rectum
\cite{poisson_will_2014}, which gives
\be
r_{\rm peri} = R \frac{v_\infty^2 b^2}{v_{\rm esc}^2 R^2} .
\ee
The condition that the star enters the cloud, $r_{\rm peri} < R$, gives an upper bound on the impact
parameter for a given relative velocity
\be
b_{\rm max} = R \frac{v_{\rm esc}}{v_\infty} \gg R , \;\;
\ecc_{\rm max}^2 = 1 + \frac{4 v_\infty^2}{v_{\rm esc}^2} \simeq 1 ,
\label{eq:bmax}
\ee
for $v_\infty \ll v_{\rm esc}$. Thus, the assumption $\ecc \simeq 1$ is self-consistent.

From Eqs.(\ref{eq:dGamma}) and (\ref{eq:bmax}), the encounter rate with clouds of relative velocity
below $v$, with $v < v_{\rm esc}$, reads
\be
\Gamma_{\rm enc}(<v) = \frac{3 \bar\rho_{\rm MW} v_{\rm esc}^2}{4\rho R}
\int_0^v \frac{dv_\infty}{v_{\infty}} f(v_\infty) ,
\ee
where we integrated over $b$ up to $b_{\rm max}(v_\infty)$.
If we neglect the velocity dispersion of the clumps, which are much more massive than the stars,
and take an isotropic Gaussian of 1D dispersion $\sigma$ for the 3D velocity distribution of the stars
(around the mean orbital velocity at some radius $r$), the distribution $f(v_{\infty})$ of the relative
velocity is the Maxwellian distribution
\be
f(v_\infty) = \frac{2}{\sqrt{2\pi} \sigma^3} v_\infty^2 e^{-v_\infty^2/(2\sigma^2)} .
\label{eq:fv-Gaussian}
\ee
This gives
\be
\Gamma_{\rm enc}(<v) = \frac{3 \bar\rho_{\rm MW} v_{\rm esc}^2}{2\sqrt{2\pi} \sigma \rho R}
\left( 1 - e^{-v^2/(2\sigma^2)} \right).
\label{eq:Gamma-enc-v}
\ee
This expression holds as long as $\sigma \lesssim v_{\rm esc}$, so that the contribution from
high-velocity stars that violate the assumption $v_{\infty} \ll v_{\rm esc}$ used in (\ref{eq:bmax})
is negligible.
This is indeed the case as $\sigma \sim 100$ km/s
\cite{2006MNRAS.369.1688D,KingIII_2015,Anguiano_2020}
whereas we obtained in (\ref{ex:v-esc})
$v_{\rm esc} = 217$ km/s.
This condition on the velocity will be even  satisfied better in the following section
where we derive the upper bound (\ref{eq:ex-vmax}) on the star velocity by taking into account
the energy loss condition.

From (\ref{eq:Gamma-enc-v}) we then obtain the total total probability of encounter with a dark matter
cloud, for a given neutron star, over a Hubble time $t_H=1/H$,
\be
P_{\rm enc} = \Gamma_{\rm enc} t_H = \frac{3 \bar\rho_{\rm MW} v_{\rm esc}^2 t_H}
{2\sqrt{2\pi} \sigma \rho R} \simeq 3 \times 10^{-3} ,
\label{eq:P-enc}
\ee
where we take $\bar\rho_{\rm MW} = 0.4$ Gev/cm$^3$ for the mean dark matter density,
$\sigma = 100$ km/s, and the cloud mass and radius from (\ref{ex:v-esc}).
It is expected that there are over 100 million neutron stars in the Milky Way
but only $10^4$ active radio pulsars \cite{Camenzind2007}.
This means that  only 30 currently active pulsars
have encountered at least once a scalar dark matter cloud.
Moreover, for a given pulsar the encounter probability is only $0.3 \%$.
As we shall see below, this probability is even further reduced when we consider the capture rate.

\subsection{Capture rate by dynamical friction}

After the star enters the dark matter cloud, it will lose some energy by dynamical friction. We assume
the Chandrasekhar expression \cite{Chandrasekhar:1943ys} for this drag force,
\be
F_{\rm dyn} = \frac{4\pi \rho {\cal G}^2 m_\star^2}{v^2} \Lambda ,
\ee
where $\Lambda$ is the Coulomb logarithm, which we will take constant.
Within an order of magnitude, this expression applies to a wide variety of regimes, from collisionless
systems when $v$ is larger than the velocity dispersion of the cloud
\cite{Chandrasekhar:1943ys,Binney2008}
to fuzzy and self-interacting scalar dark matter when the velocity is larger than the local speed
of sound \cite{Hui:2016ltb,Berezhiani:2019pzd,Boudon:2023qbu}.
The energy loss as the star moves inside the cloud is $\Delta E = \int F_{\rm dyn} d\ell$,
which we estimate as
\be
\Delta E =  \frac{4\pi \rho {\cal G}^2 m_\star^2}{v_{\rm esc}^2} \Lambda R .
\ee
Here we used that the star enters the cloud with a velocity $v \simeq v_{\rm esc}$ because
$v_\infty \ll v_{\rm esc}$, as we check below.
To become bound to the cloud, the star must lose enough energy, $\Delta E > E_{\rm init}$,
where $E_{\rm init} = m_\star v_\infty^2/2$ is its initial energy.
This gives the condition
\be
v_\infty < v_{\rm max} \;\; \mbox{with} \;\;
v_{\rm max}^2 = v_{\rm esc}^2 \frac{3\Lambda}{2} \frac{m_\star}{M} \ll v_{\rm esc}^2 .
\label{eq:vmax}
\ee
Thus, we find indeed $v_{\infty} \ll v_{\rm esc}$ because the mass of the star is much smaller than the
mass of the cloud, $m_\star \ll M$.
For dark matter clouds such as (\ref{ex:v-esc}) and a neutron star of mass $m_\star = 2 M_\odot$,
we obtain for $\Lambda=1$
\be
m_\star = 2 M_\odot : \;\; v_{\rm max} = 0.5 \, {\rm km/s}.
\label{eq:ex-vmax}
\ee
This small value of $v_{\rm max}$ means that dynamical friction is quite inefficient to dissipate
the kinetic energy of the star in one crossing of the cloud.
Then, the capture rate is $\Gamma_{\rm cap} = \Gamma_{\rm enc}(<v_{\rm max})$, with the maximum
velocity (\ref{eq:vmax}).
From Eq.(\ref{eq:Gamma-enc-v}), with $v_{\rm max} \ll \sigma$, we obtain
\be
\Gamma_{\rm cap} = \frac{3 \sqrt{2\pi} \Lambda {\cal G}^2 \bar\rho_{\rm MW} m_\star}
{\sigma^3} ,
\ee
which happens to be independent of the mass and density of the dark matter clouds
(in the regime of validity).
Over the Hubble time $t_H=1/H$, this gives the probability of capture for a given neutron star
\be
P_{\rm cap} = \Gamma_{\rm cap} t_H \simeq 4 \times 10^{-8} ,
\label{eq:P-cap}
\ee
where we take $\bar\rho_{\rm MW} = 0.4$ Gev/cm$^3$ for the mean dark matter density,
$m_\star = 2 M_\odot$, $\sigma = 100$ km/s and $\Lambda=1$.
Therefore, a few neutron stars may be captured in the Milky Way, but the probability
that a given pulsar is embedded in a dark matter cloud is extremely low.
The relatively inefficient dissipation by dynamical friction yields a much reduced capture
probability as compared with the encounter probability (\ref{eq:P-enc}).

\subsection{Random location}

In the previous sections we have derived the encounter and capture probabilities of distant stars.
If the dark matter clouds cover a sufficiently large fraction of the Milky Way volume,
they will contain neutron stars that simply happen to be located within the clouds.
Assuming random and uncorrelated distributions of the stars and of the clouds, the volume fraction
occupied by the clumps is $V_{\rm occ}/V_{\rm tot} = \bar\rho_{\rm MW}/\rho$.
This gives a probability to be located within a dark matter cloud
\be
P_{\rm loc} = \frac{\bar\rho_{\rm MW}}{\rho} \simeq 10^{-10} ,
\ee
for the cloud inner density (\ref{eq:ex-rho}).
This is even smaller than (\ref{eq:P-cap}), so that most stars that are located within dark matter
clouds were captured by dynamical friction (or other more efficient processes).

\bibliography{ref}

\end{document}